\documentstyle[12pt]{article}
\newlength{\mathspace}
\def\lb{\lbrack}
\def\rb{\rbrack}
\def\({\left(}
\def\){\right)}
\def\pa{\partial}
\newcommand\DB{{Darboux-B\"{a}cklund}~}

\begin{document}
\baselineskip=0.7cm
\setlength{\mathspace}{2.5mm}



\begin{titlepage}

    \begin{normalsize}
     \begin{flushright}
                
                 solv-int/9710026\\
     \end{flushright}
    \end{normalsize}
    \begin{large}
       \vspace{1cm}
       \begin{center}
         {\bf The sAKNS Hierarchy} 
       \end{center}
    \end{large}

  \vspace{5mm}

\begin{center}
           
             \vspace{.5cm}

            Henrik Aratyn
           \footnote{E-mail address: aratyn@uic.edu}\\
Department of Physics\\
 
University of Illinois at Chicago\\
Chicago, IL 60607, USA\\
and \\
Ashok Das\\
Department of Physics and Astronomy\\
University of Rochester\\
Rochester, N.Y. 14627, USA\\

      \vspace{2.5cm}

    \begin{large} ABSTRACT \end{large}
        \par
\end{center}
 \begin{normalsize}
\ \ \ \
We study, systematically, the properties of the supersymmetric AKNS (sAKNS)
hierarchy. In particular, we discuss the Lax representation in terms of a
bosonic Lax operator and  some special features of the equations and construct
the bosonic local charges as well as the fermionic nonlocal charges associated 
with the
system starting from the Lax operator. We obtain the Hamiltonian structures of
the system and check  the Jacobi identity through the method of prolongation.
We also show that this hierarchy of equations can equivalently be described in
terms of a fermionic Lax operator. We obtain the zero curvature formulation as
well as the conserved charges of the system starting from this fermionic Lax
operator which suggests a connection between the two. Finally, starting from
the fermionic description of the system, we construct the soliton solutions for
this system of equations through Darboux-B\"acklund transformations and        
describe some open problems.
\end{normalsize}

\end{titlepage}
\vfil\eject

\begin{large}
\noindent{\bf 1. Introduction:}
\end{large}
\vspace{.5cm}

The AKNS hierarchy[1-3] is an important bosonic, integrable hierarchy which has
played a fundamental role in the development of many interesting ideas. This
hierarchy, among other things, includes all the well known integrable models
such as the KdV equation, the mKdV equation, the nonlinear Schr\"odinger
equation, sine-Gordon equation. In its original formulation, the AKNS
hierarchy was described in terms of a matrix Lax operator
\begin{equation}
L = \frac{\partial}{\partial x} - q \sigma_{+} - r \sigma_{-} + i\zeta
\sigma_{3}\label{I_{1}}
\end{equation}
where $q$ and $r$ represent the dynamical variables while $\zeta$ represents
the spectral parameter. Furthermore, the $\sigma$'s represent the $2\times 2$
Pauli matrices and, in particular, $\sigma_{\pm} = \frac{1}{2} (\sigma_{1}\pm
i\sigma_{2})$. Different integrable hierarchies would result from this Lax
operator with different identifications of the dynamical variables. Thus, for
example, $r\sim\psi$ and $q\sim\psi^{*}$ would lead to the hierarchy associated
with the nonlinear Schr\"odinger equation.

The AKNS hierarchy can also be described in terms of  the more conventional    
scalar Lax operator of the form[4-5]
\begin{equation}
L = \partial + q \partial^{-1} r\label{I_{2}}
\end{equation}
which can also be seen to follow from the linear equation associated with the
matrix Lax operator in eq. (\ref{I_{1}}). This way of describing the AKNS
hierarchy, however, has led to further interesting properties associated with
the system. Among other things, it has led to a new understanding of the
gradings associated with the zero curvature formulation of the integrable
models.

In recent years, there has been a lot of interest in understanding the
properties of supersymmetric integrable models[6]. Several such models have
already been constructed[7-9] and more recently, a supersymmetric formulation  
of the
AKNS (sAKNS) hierarchy has also been given[10-11]. It is expected to be at     
least as   
important in the study of supersymmetric integrable models as the AKNS 
hierarchy is within the context of the bosonic integrable systems. Therefore,
it is worth studying the properties of the sAKNS hierarchy systematically which
we carry out in this paper. In section 2, we describe the sAKNS hierarchy in
terms of the conventional bosonic Lax operator in superspace
giving the explicit form of the dynamical equations upto the first few orders.
We construct the local, bosonic conserved charges as well as the nonlocal,
fermionic conserved charges associated with the system starting from the Lax
operator. We also review very briefly the zero curvature formulation of this   
system[12-13] in this section. In section 3, we construct the Hamiltonian      
structures
associated with this system pointing out some of the subtleties. We check
Jacobi identity for these structures using the method of prolongation[14-15].  
We also
show that the Hamiltonian structures are compatible making it a genuinely
bi-Hamiltonian system. We also construct, in this section, the recursion       
operator, with a vanishing Nijenhuis torsion tensor, which relates  the        
different Hamiltonians of the theory. In section 4, we show that the sAKNS
hierarchy can also be described in terms of a fermionic Lax operator. This is,
in fact, 
the only integrable system we know of which allows a description in terms of a
bosonic as well as a fermionic Lax operator. We construct the local conserved  
charges
and the zero curvature formulation starting with this fermionic Lax operator.
However, the construction of the nonlocal, fermionic conserved charges, from
the fermionic Lax operator, as well
as a transformation relating the two Lax operators directly remains an open
question. In section 5, we construct soliton solutions associated with this
system starting from the fermionic Lax operator through a Darboux-B\"acklund
transformation and compare its properties with those
of such solutions obtained from an algebraic dressing method[13]. Finally, we  
close with some conclusions in section 6.
\vspace{.5cm}

\begin{large}
\noindent{\bf 2. The sAKNS System:}
\end{large}

\vspace{.5cm}
The supersymmetric AKNS hierarchy can be described by the bosonic Lax operator
\begin{equation}
L = D^{2} + \phi D^{-1} \psi\label{A_{1}}
\end{equation}
where the supercovariant derivative is defined to be
\begin{equation}
D = \frac{\partial}{\partial \theta} + \theta \frac{\partial}{\partial
x}\label{A_{2}}
\end{equation}
Here $\theta$ is the Grassmann coordinate of the superspace and $\phi$ and
$\psi$ represent a bosonic and a fermionic superfield respectively. It is also
straightforward to see that $D^{2} = \partial_{x}$. The
hierarchy of equations (the sAKNS hierarchy) can be obtained from the scalar
Lax equation
\begin{equation}
\frac{\partial L}{\partial t_{n}} = \left[(L^{n})_{+}, L\right]\label{A_{3}}
\end{equation}
where $n = 0,1,2,\cdots$ represents the flow of the hierarchy. We note here the
first few flows of the hierarchy, namely,
\vfil\eject
\begin{eqnarray}
\frac{\partial \phi}{\partial t_{0}} & = & \phi\nonumber\\
\frac{\partial \psi}{\partial t_{0}} & = & - \psi\label{a_{4}}\\
\frac{\partial \phi}{\partial t_{1}} & = & (D^{2} \phi)\nonumber\\
\frac{\partial \psi}{\partial t_{1}} & = & (D^{2} \psi)\label{A_{4}}\\
\frac{\partial \phi}{\partial t_{2}} & = & (D^{4} \phi) + 2 \phi (D \phi
\psi)\nonumber\\
\frac{\partial \psi}{\partial t_{2}} & = & - (D^{4} \psi) - 2 \psi (D \phi
\psi)\label{A_{5}}\\
\frac{\partial \phi}{\partial t_{3}} & = & (D^{6} \phi) + 3 \phi (D
(D^{2}\phi)\psi) + 3 (D^{2}\phi) (D\phi\psi)\nonumber\\
\frac{\partial \psi}{\partial t_{3}} & = & (D^{6} \psi) + 3 \phi (D^{2} \psi
(D\psi))\label{A_{6}}
\end{eqnarray}
and so on. It is worth noting here a peculiarity of this hierarchy of
equations, namely, the lowest order equation, which arises from $(L^{0})_{+}=1$
is nontrivial mainly because the Lax operator in eq. (\ref{A_{1}}) is composite
in the dynamical variables as opposed to the usual cases where the coefficients
of the pseudodifferential operators are linear in the dynamical variables
leading to the fact that nontrivial flows exist only for $n= 1,2,\cdots$.

Given the Lax operator in eq. (\ref{A_{1}}), we can immediately construct the
local conserved quantities of the system. They are simply given by
\begin{equation}
H_{n} = - \frac{1}{n} sTr L^{n} = - \frac{1}{n} \int dz\; sRes L^{n}\label{A_{7}}
\end{equation}
where $z$ represents the coordinates of the superspace and the superTrace is
defined as the integral over the superspace of the superResidue which
corresponds to the coefficient of the $D^{-1}$ term of the pseudodifferential
operator. Because $L$ is bosonic, it is clear from eq. (\ref{A_{7}}), that
these conserved quantities are bosonic as well. The first few of these
conserved quantities have the explicit form
\begin{eqnarray}
H_{1} & = & \int dz\; \phi \psi\nonumber\\
H_{2} & = & \int dz\; (D^{2}\phi) \psi\nonumber\\
H_{3} & = & \int dz\; \left[(D^{4}\phi) \psi + \phi^{2} \psi
(D\psi)\right]\nonumber\\
H_{4} & = & \int dz\; \left[(D^{6}\phi) \psi + 3 \phi \psi (D\psi)
(D^{2}\phi)\right]\label{A_{8}}
\end{eqnarray}
These quantities can be easily seen to be conserved from the form of the Lax
equation in eq. (\ref{A_{2}}) or by explicit computation.

In addition to the local conserved charges, the theory does contain conserved,
nonlocal fermionic charges which can also be obtained from the Lax operator as
\begin{equation}
Q_{n} = (-1)^{n+1} sTr L^{(n+\frac{1}{2})}\label{A_{9}}
\end{equation}
It is clear from the definition of these charges that they are fermionic since
the Lax operator is bosonic and the explicit form of the first two is given by
\begin{eqnarray}
Q_{0} & = & \int dz\; (D^{-1} \phi \psi)\nonumber\\
Q_{1} & = & \int dz\; \left[\phi (D\psi) + \frac{1}{2} (D^{-1} \phi \psi)^{2} -
(D^{-1} (D^{2}\phi) \psi) + 2 \phi \psi (D^{-2} \phi \psi)\right]\label{A_{10}}
\end{eqnarray}
It is clear that they are manifestly fermionic and nonlocal. The fact that
these charges are conserved is not obvious from the Lax equation. However, a
little bit of algebra (tedious) shows that they are, indeed, conserved under
the flows of the hierarchy in eqs. (\ref{A_{3}})--(\ref{A_{6}}). The theory is
supersymmetric and so, we expect one of these fermionic charges to generate the
supersymmetry transformations and this depends on the particular Hamiltonian
structure used.

For the sake of completeness as well as for later use, we discuss  here, 
briefly,the zero curvature formulation of this hierarchy[12-13]. The Lax       
operator of eq. (\ref{A_{1}}) leads to the linear equation
\begin{equation}
(D^{2} + \phi D^{-1} \psi)\chi = \lambda \chi\label{A_{11}}
\end{equation}
Here $\lambda$ is the spectral parameter of the theory. Starting with this, as 
discussed in [12], we can obtain an $OSp(2|2)$ ($SL(2|1)$) valued potential, 
${\cal A}_{1}$,
\begin{equation}
{\cal A}_{1} = \left(\begin{array}{ccc}
\lambda & -\phi & 0\\
(D\psi) & 0 & -\psi\\
-\phi\psi & -(D\phi) & \lambda
\end{array}\right)\label{A_{12}}
\end{equation}
such that we can write the linear equation of (\ref{A_{11}}) also as a matrix
equation
\begin{equation}
\partial_{x}\left(\begin{array}{c}
\chi\\
(D^{-1}\psi\chi)\\
(D\chi)
\end{array}\right) = {\cal A}_{1}\left(\begin{array}{c}
\chi\\
(D^{-1}\psi\chi)\\
(D\chi)
\end{array}\right)\label{A_{13}}
\end{equation}
It is, then, easy to show that if we define another $OSP(2|2)$ ($SL(2|1)$) 
valued potential, ${\cal A}_{0}$ as
\begin{equation}
{\cal A}_{0} = \left(\begin{array}{ccc}
A & B & C\\
E + C\psi & F & G\\
H + \lambda C & J - \phi C & A + F
\end{array}\right)\label{A_{14}}
\end{equation}
then, the zero curvature condition associated with these potentials, namely,
\[
\partial_{t} {\cal A}_{1} - \partial_{x} {\cal A}_{0} - [{\cal A}_{0} , {\cal
A}_{1}] = 0
\]
leads to the sAKNS hierarchy of equations provided
\begin{eqnarray}
E & = & - (DG)\nonumber\\
F & = & (DC)\nonumber\\
H & = & (DA) + B \psi\nonumber\\
J & = & (DB)\nonumber\\
C_{x} & = & - G\phi + B \psi\nonumber\\
A_{x} & = & \phi (DG) - B (D\psi) + 2 C \phi \psi\label{A_{15}}
\end{eqnarray}
with the dynamical equations given by
\begin{eqnarray}
\phi_{t} & = & - [B_{x} - \phi A + (D\phi C) - \lambda B]\nonumber\\
\psi_{t} & = & - [G_{x} + \psi A - (D\psi) C + \lambda G]\label{A_{16}}
\end{eqnarray}
\vspace{.5cm}
\begin{large}
\noindent{\bf 3. Hamiltonian Structures:}
\end{large}

\vspace{.5cm}
In this section, we describe the Hamiltonian structures of the theory. Our goal
is to find the Hamiltonian structure ${\cal D}$ such that the dynamical
equations of the hierarchy can be written as
\begin{equation}
\partial_{t}\left(\begin{array}{c}
\phi\\
\psi
\end{array}\right) = {\cal D} \left(\begin{array}{c}
\frac{\delta H}{\delta \phi}\\
\frac{\delta H}{\delta \psi}
\end{array}\right)\label{B_{1}}
\end{equation}
where $H$ represents the appropriate Hamiltonian for the flow. From the form of
the dynamical equations in (\ref{A_{4}})--(\ref{A_{6}}) as well as the
forms of the conserved quantities in eq. (\ref{A_{9}}), it is straightforward
to write down the simplest of the Hamiltonian structures, namely, we see that
with
\begin{equation}
{\cal D}_{1} = \left(\begin{array}{rr}
0 & 1\\
-1 & 0
\end{array}\right)\label{B_{2}}
\end{equation}
we can write all the equations of the hierarchy as
\begin{equation}
\partial_{t_{n}} \left(\begin{array}{c}
\phi\\
\psi
\end{array}\right) = {\cal D}_{1} \left(\begin{array}{c}
\frac{\delta H_{n}}{\delta \phi}\\
\frac{\delta H_{n}}{\delta \psi}
\end{array}\right)\label{B_{3}}
\end{equation}
The structure ${\cal D}_{1}$ is a simple structure with constant elements and,
therefore, trivially satisfies the Jacobi identity. Thus, we can identify this
with the first Hamiltonian structure of the hierarchy.

However, as we know, integrable hierarchies have, in general, several distinct
Hamiltonian structures. Finding the higher ones, though, can be tricky as was
already noticed in [9, 16]. Let us point out how one can run into problems here
also. With some algebra, one can show that the structure
\begin{equation}
\tilde{\cal D}_{2} = \left(\begin{array}{cc}
-2 \phi D^{-1} \phi & D^{2} + 2 \phi D^{-1} \psi\\
D^{2} + 2 \psi D^{-1} \phi & - 2 \psi D^{-1} \psi
\end{array}\right)\label{B_{4}}
\end{equation}
gives equations (\ref{A_{5}})--(\ref{A_{6}}) through the relation (for $n=0,1$)
\begin{equation}
\partial_{t_{n+1}} \left(\begin{array}{c}
\phi\\
\psi
\end{array}\right) = \tilde{\cal D}_{2} \left(\begin{array}{c}
\frac{\delta H_{n}}{\delta \phi}\\
\frac{\delta H_{n}}{\delta \psi}
\end{array}\right)
\end{equation}
This is, however, a much more complicated structure than ${\cal D}_{1}$ and
Jacobi identity is not obvious from the form of this structure. A careful
calculation shows that $\tilde{\cal D}_{2}$ does not satisfy the Jacobi identity and,
therefore , cannot represent a true Hamiltonian structure of the theory. In
fact, if we try to derive eq. (\ref{A_{7}}) with this structure, it fails
showing that the structure is not quite complete. It misses some terms which
give a vanishing contribution for the low order equations and, therefore, do
not make their presence manifest.

The true second Hamiltonian structure of the hierarchy can be determined with
some work to be
\begin{equation}
{\cal D}_{2} = \left(\begin{array}{ccc}
\begin{array}{c}
-\phi D^{-2}\phi D - D \phi D^{-2}\phi\\ 
- 2\phi D^{-2}\phi\psi D^{-2}\phi\\
 
\end{array} & \begin{array}{c}
D^{2} + D\phi D^{-2}\psi + \phi D^{-2}(D\psi)\\ 
+ 2\phi D^{-2}\phi\psi D^{-2}\psi\\

\end{array}\\
\begin{array}{c}
D^{2} + \psi D^{-2}\phi D + (D\psi) D^{-2}\phi\\ 
+ 2\psi D^{-2}\phi\psi D^{-2}\phi
\end{array} & \begin{array}{c} 
-\psi D^{-2}(D\psi) - (D\psi)D^{-2}\psi\\ 
- 2\psi D^{-2}\phi\psi D^{-2}\psi
\end{array}
\end{array}\right)\label{B_{5}}
\end{equation}
It is now easy to check that the equations of the hierarchy (except for the
lowest one) can now be written as ($n=0,1,2,\cdots$)
\begin{equation}
\partial_{t_{n+1}} \left(\begin{array}{c}
\phi\\
\psi
\end{array}\right) = {\cal D}_{2} \left(\begin{array}{c}
\frac{\delta H_{n}}{\delta \phi}\\
\frac{\delta H_{n}}{\delta \psi}
\end{array}\right)\label{B_{6}}
\end{equation}

This is, of course, a much more complicated structure. However, the Jacobi
identity can be verified through the method of prolongation which we describe
briefly. Let us consider a graded matrix one form $\Omega$ defined as
\begin{equation}
\Omega = \left(\begin{array}{c}
\Omega_{f}\\
\Omega_{b}
\end{array}\right)\label{B_{7}}
\end{equation}
where $\Omega_{b}$ and $\Omega_{f}$ represent respectively bosonic and
fermionic components of the matrix one form. Given this and the structure
${\cal D}_{2}$, we can now construct the bivector associated with the structure
${\cal D}_{2}$ as
\begin{eqnarray}
\Theta_{{\cal D}_{2}} & = & \frac{1}{2} \int dz\; ({\cal
D}_{2}\Omega)_{\alpha} \wedge \Omega_{\alpha}\nonumber\\
 & = & \int dz \left[(D^{2}\Omega_{b})\wedge \Omega_{f} + (\phi\Omega_{f} -
\psi\Omega_{b})\wedge\right.\nonumber\\
 &  & \left.\left\{D^{-2}(\phi(D\Omega_{f}) - (D\psi)\Omega_{b} +
\phi\psi(D^{-2}(\phi\Omega_{f}-\psi\Omega_{b})))\right\}\right]\label{B_{8}}
\end{eqnarray}
Here $\alpha = 1,2$ runs over the two components of the matrix and we have used
integration by parts. The structure ${\cal D}_{2}$ can be shown to satisfy
Jacobi identity provided the prolongation of this bivector vanishes.

The prolongation can be calculated by noting that the prolongations for the
basic variables are defined to be (see [14-15] for details)
\begin{equation}
{\bf pr\;v}_{{\cal D}_{2}\Omega}\left(\begin{array}{c}
\phi\\
\psi
\end{array}\right) = \left(\begin{array}{c}
({\cal D}_{2}\Omega)_{1}\\
({\cal D}_{2}\Omega)_{2}
\end{array}\right)\label{B_{9}}
\end{equation}
With some tedious algebra and using eq. (\ref{B_{9}}), we can show that (upto
surface terms)
\begin{equation}
{\bf pr\;v}_{{\cal D}_{2}\Omega} (\Theta_{{\cal D}_{2}}) = 0\label{B_{10}}
\end{equation}
This shows that ${\cal D}_{2}$ satisfies the Jacobi identity and, indeed,
represents the true second Hamiltonian of the theory. Furthermore, it is also
easy to show that the structures ${\cal D}_{1}$ and ${\cal D}_{2}$ are
compatible. Namely, let
\begin{equation}
{\cal D} = {\cal D}_{2} + \alpha {\cal D}_{1}\label{b_{9}}
\end{equation}
where $\alpha$ is an arbitrary constant parameter. Then, from the simple
structure of ${\cal D}_{1}$ in eq. (\ref{B_{2}}), it is straightforward to
check that
\begin{equation}
{\bf pr\;v}_{{\cal D}\Omega}(\Theta_{{\cal D}}) = 0
\end{equation}
showing that any linear combination of ${\cal D}_{1}$ and ${\cal D}_{2}$ is
also a Hamiltonian structure. 

The compatibility of ${\cal D}_{1}$ and ${\cal D}_{2}$, suggests that we can  
define the recursion operator for the theory which corresponds to
\begin{eqnarray}
{\cal R} & = & {\cal D}_{1}^{-1} {\cal D}_{2}\nonumber\\
 & = & \left(\begin{array}{cc}
\begin{array}{c}
-D^{2} - \psi D^{-2}\phi D - (D\psi)D^{-2}\phi\\ 
- 2\psi D^{-2}\phi\psi D^{-2}\phi\\

\end{array} & \begin{array}{c} 
\psi D^{-2}(D\psi) + (D\psi)D^{-2}\psi\\ 
+ 2\psi D^{-2}\phi\psi D^{-2}\psi\\

\end{array}\\
\begin{array}{c}
-\phi D^{-2}\phi D - D\phi D^{-2}\phi\\ 
- 2\phi D^{-2}\phi\psi D^{-2}\phi 
\end{array} & \begin{array}{c}
D^{2} + D\phi D^{-2}\psi + \phi D^{-2}(D\psi)\\ 
+ 2\phi D^{-2}\phi\psi D^{-2}\psi
\end{array}
\end{array}\right)\label{B_{11}}
\end{eqnarray}
which will give the recursion relation between the Hamiltonians as
\begin{equation}
\left(\begin{array}{c}
\frac{\delta H_{n+1}}{\delta \phi}\\
\frac{\delta H_{n+1}}{\delta \psi}
\end{array}\right) = {\cal R}\left(\begin{array}{c}
\frac{\delta H_{n}}{\delta \phi}\\
\frac{\delta H_{n}}{\delta \psi}
\end{array}\right)\label{B_{12}}
\end{equation}
as well as the higher order Hamiltonian structures, for example,
\[
{\cal D}_{3} = {\cal D}_{2} {\cal R}
\]
and so on.  Furthermore, the Nijenhuis torsion tensor associated with this
recursion operator would vanish implying again the integrability of this       
hierarchy of equations.

It is also worth pointing out here that we could try to obtain an alternate
supersymmetry hierarchy associated with the AKNS hierarchy as follows. If we
choose as the Lax operator
\begin{equation}
L = D^{2} + \phi D^{-2} (D\psi)\label{B_{13}}
\end{equation}
which differs only slightly from the Lax operator in eq. (\ref{A_{1}}), the
equations
\begin{equation}
\frac{\partial L}{\partial t_{n}} = \left[(L^{n})_{+} , L\right]\label{B_{14}}
\end{equation}
would correspond to the sAKNS-B hierarchy which is an alternate
supersymmetrization[17] of the AKNS hierarchy. Thus, for example, the second   
order equation following from eq. (\ref {B_{14}}) would be
\begin{eqnarray}
\frac{\partial \phi}{\partial t} & = & (D^{4}\phi) + 2 \phi^{2}
(D\psi)\nonumber\\
\frac{\partial \psi}{\partial t} & = & -(D^{4}\psi) - 2 (D^{-1} \phi
(D\psi)^{2})\label{B_{15}}
\end{eqnarray}
which can be compared with eq. (\ref{A_{5}}). The second equation of
(\ref{B_{15}}), however, is nonlocal, as was also noted in [18] and so, the
alternate supersymmetrization for the case of AKNS hierarchy does not appear to
be particularly useful.

We also note from the form of the Lax operator in (\ref{A_{1}}) that the
terms are not linear in the dynamical variables. Therefore, it is not clear how
one can define a dual which will give rise to a linear functional of the
dynamical variables. Consequently, the question of obtaining the Hamiltonian
structures of the theory from the Gelfand-Dikii brackets or through the
R-matrix approach remains open.
\vfil\eject

\begin{large}
\noindent{\bf 4. A Fermionic Lax Description:}
\end{large}

\vspace{.5cm}
The sAKNS hierarchy discussed in the above sections is quite peculiar in that,
in addition to its description in terms of a bosonic Lax operator in eq. 
(\ref{A_{1}}, it can also be described equivalently by a fermionic Lax
operator. Let us consider the Lax operator
\begin{equation}
\Lambda = D + \phi D^{-2} \psi\label{C_{1}}
\end{equation}
where $\phi$ and $\psi$ denote the superfields discussed in the earlier
sections. It is clear that this Lax operator, unlike the one in eq.
(\ref{A_{1}}), is fermionic. Furthermore, it is straightforward to check that
the Lax equation ($n=0,1,2,\cdots$)
\begin{equation}
\frac{\partial \Lambda}{\partial t_{n}} = \left[(\Lambda^{2n})_{+} ,
\Lambda\right]\label{C_{2}}
\end{equation}
gives the sAKNS hierarchy of equations in eq. (\ref{A_{2}}) or more explicitly
in eqs. (\ref{a_{4}})--(\ref{A_{6}}). This is, therefore, the only hierarchy of
equations that we know of which can be described in terms of a bosonic as well
as a
fermionic Lax operator. Furthermore, there is no obvious relation between the
two Lax operators in eqs. (\ref{A_{1}}) and (\ref{C_{1}}) although we will
describe some equivalences later in this section. We note here that the odd
powers of $\Lambda$, namely, $(\Lambda^{2n+1})_{+}$ in eq. (\ref{C_{2}}) do not
lead to any consistent equation (with anti-commutation relation, of course).

Since this is a new Lax operator, it is worth studying its properties
systematically. First, let us look at the conserved quantities following from
this Lax operator. It is straightforward to check that
\begin{equation}
sTr (\Lambda)^{n} = 0\hspace{.5in}{\rm for~ any}\;n\label{C_{3}}
\end{equation}
On the other hand, we have checked explicitly upto the fourth term and
conjecture that
\begin{equation}
Res \Lambda^{2n} = - (D sRes L^{n})\label{C_{4}}
\end{equation}
so that the local conserved quantities in eq. (\ref{A_{7}}) can equivalently be
written as
\begin{equation}
H_{n} = \frac{1}{n} \int dz\; (D^{-1} Res \Lambda^{2n})\label{C_{5}}
\end{equation}
Since $\Lambda$ is a fermionic operator, it would appear that the fermionic
conserved quantities should follow in a simple manner. However, we have not
succeeded in obtaining the fermionic, nonlocal conserved quantities of eq.
(\ref{A_{9}})--(\ref{A_{10}}) from $\Lambda$. That remains an open question.

It would be interesting if we can relate the two Lax operators in some way so
that the calculations with either of them would simplify. However, we have not,
so far, found a transformation that would directly relate the two except for
the observation that, under
\begin{equation}
\phi\rightarrow D^{-1} \phi D\label{C_{6}}
\end{equation}
we have
\begin{eqnarray}
\Lambda & \rightarrow & D + D^{-1} \phi D^{-1} \psi\nonumber\\
        & = & D^{-1} (D^{2} + \phi D^{-1} \psi)\nonumber\\
        & = & D^{-1} L\label{C_{7}}
\end{eqnarray}
This is, however, not very useful in practical calculations. We note here that
we can write
\begin{eqnarray}
\Lambda^{2} & = & D^{2} + D\phi D^{-2}\psi + \phi D^{-2}\psi D + \phi
D^{-2}\phi\psi D^{-2}\psi\nonumber\\
 & = & D^{2} + \bar{\phi} D^{-2} \psi - \phi D^{-2} \bar{\psi}\label{C_{8}}
\end{eqnarray}
where
\begin{eqnarray}
\bar{\phi} & = & (D\phi) + (D^{-2} \phi \psi) \phi\nonumber\\
\bar{\psi} & = & - (D\psi) + (D^{-2} \phi\psi) \psi\label{C_{9}}
\end{eqnarray}
are the variables already defined in [12] in connection with a Hamiltonian
description of this system. This way of defining $\Lambda^{2}$ simplifies the
calculation of the local conserved quantities of the theory greatly.

Let us next note that if we start with the Lax operator in eq. (\ref{C_{1}}),
we can write a linear equation of the form
\begin{equation}
\Lambda \chi = (D + \phi D^{-2} \psi)\chi = 0\label{C_{10}}
\end{equation}
{}From eq. (\ref{C_{10}}), we can write the matrix equation
\begin{equation}
D \left(\begin{array}{c}
(D^{-1}\psi \chi)\\
\chi\\
(D^{-2} \psi \chi)
\end{array}\right) = \left(\begin{array}{ccr}
0 & \psi & 0\\
0 & 0 & -\phi\\
1 & 0 & 0
\end{array}\right)\label{C_{11}}
\end{equation}
It now follows from this, as well as the properties of the supercovariant
derivative, that
\begin{equation}
\partial_{x}\left(\begin{array}{c}
(D^{-1}\psi \chi)\\
\chi\\
(D^{-2} \psi \chi)
\end{array}\right) = {\cal A}_{1} \left(\begin{array}{c}
(D^{-1}\psi \chi)\\
\chi\\
(D^{-2} \psi \chi)
\end{array}\right)\label{C_{12}}
\end{equation}
where
\begin{equation}
{\cal A}_{1} = \left(\begin{array}{rcc}
0 & (D\psi) & \phi\psi\\
-\phi & 0 & -(D\phi)\\
0 & \psi & 0
\end{array}\right)\label{C_{13}}
\end{equation}
Furthermore, if we define a graded matrix
\begin{equation}
{\cal A}_{0} = \left(\begin{array}{ccc}
A & E - C\psi & H\\
B & F & J + \phi C\\
C & G & A+F
\end{array}\right)\label{C_{14}}
\end{equation}
then, it is straightforward to see that the zero curvature condition
\[
\partial_{t} {\cal A}_{1} - \partial_{x} {\cal A}_{0} - [{\cal A}_{0} , {\cal
A}_{1}] = 0
\]
leads to the sAKNS hierarchy of equations provided
\begin{eqnarray}
F & = & - (DC)\nonumber\\
E & = & (DG)\nonumber\\
J & = & (DB)\nonumber\\
H & = & (DA) - B\psi\nonumber\\
A_{x} & = & (DG) \phi + B (D\psi) - 2 C \phi\psi\nonumber\\
C_{x} & = & G\phi + B\psi\label{C_{15}}
\end{eqnarray}
It is easy now to check from eqs. (\ref{A_{12}})--(\ref{A_{15}}) that these are
the same as before with proper identification (and $\lambda = 0$). This is an 
alternate way to see
that the fermionic Lax operator gives rise to the entire hierarchy of sAKNS
equations.

An alternate way of establishing this equivalence with the spectral parameter  
is to note that if we start with the linear matrix equation
\begin{equation}  
{\widetilde \Lambda} \left(\begin{array}{c}
\chi_{b}\\
\chi_{f}
\end{array}\right) = \sqrt{\lambda}\left(\begin{array}{cc}
0 & 1\\
1 & 0
\end{array}\right) \left(\begin{array}{c}
\chi_{b}\\
\chi_{f}
\end{array}\right)\label{C_{16}}
\end{equation}
where ${\widetilde \Lambda}$ is obtained from $\Lambda$ by conjugation
\begin{equation}  
{\widetilde \Lambda}= - \Lambda^{\dagger} = D + \psi D^{-2} \phi 
\label{lambdap}
\end{equation}
and $\chi_{b}$ and $\chi_{f}$ are two bosonic and fermionic superfield wave
functions respectively,
then, this set of two coupled equations is equivalent to the  two uncoupled
equations of the form
\begin{equation}
{\widetilde \Lambda}^{2} \chi = 
(D^{2} - \bar{\psi} D^{-2}\phi - \psi D^{-2} \bar{\phi})\chi
= \lambda \chi\label{C_{17}}
\end{equation}
It is now easy to obtain from this that
\begin{equation}
\partial_{x}\left(\begin{array}{c}
e^{( - \lambda x)} (D^{-2}\phi\chi)\\
e^{( - \lambda x)} \chi\\
e^{( - \lambda x)} (D^{-2}{\bar\psi} \chi)
\end{array}\right) = \left(\begin{array}{ccr}
-\lambda  & \phi & 0\\
\bar{\psi} & 0 & \psi\\
0 & \bar{\phi} & - \lambda 
\end{array}\right) 
\left(\begin{array}{c}
e^{( - \lambda x)} (D^{-2}\phi\chi)\\
e^{( -\lambda x)} \chi\\
e^{( - \lambda x)} (D^{-2}{\bar\psi} \chi)
\end{array}\right)\label{C_{18}}
\end{equation}
This is the same as the matrix obtained earlier in [12] in connection with the
conventional bosonic Lax operator. This establishes a
connection between the matrix eigenvalue problems of two descriptions 
although a more direct connection remains an open question. 
\vspace{.5cm}

\begin{large}
\noindent{\bf 5. Soliton Solutions:}
\end{large}

\vspace{.5cm}
The soliton solutions for the sAKNS hierarchy have already been constructed
earlier [13] from the conventional, bosonic Lax operator through the algebraic
dressing method. Here,
we would like to discuss the construction of soliton solutions for the
fermionic Lax operator
${\widetilde \Lambda}$ as elements on the Darboux-B\"{a}cklund orbit.
Let us consider a general Darboux-B\"{a}cklund transformation generated by
\begin{equation}
T = \xi \pa \xi^{-1} = \xi D^{2} \xi^{-1}
\label{T}
\end{equation}
where $\xi$ is an arbitrary bosonic superfield. From this, we obtain
\begin{equation}
T D T^{-1} = D + \xi \(D^3 \ln \xi \) D^{-2} \xi^{-1}
\label{tdt}
\end{equation}
Similarly, we can also show that
\begin{equation}
T \psi D^{-2} \phi T^{-1} =  \xi \( \xi^{-1} \psi \int (\phi \xi ) \)^{\prime}
D^{-2} \xi^{-1} - \xi \(\xi^{-1} \psi  \)^{\prime}
D^{-2}\(\int (\phi \xi )\) \xi^{-1} 
\label{tpdpt}
\end{equation}
We want the sum of the terms on right hand sides of (\ref{tdt}) and
(\ref{tpdpt}) representing
$ T {\widetilde \Lambda} T^{-1}$ to have the same form as
${\widetilde \Lambda} = D + \psi D^{-2} \phi$
in (\ref{lambdap}).

It is clear that there are only two possible solutions.
First, if $\xi$  satisfies $\(\xi^{-1} \psi  \)^{\prime}=0$
or $\xi^{-1} \psi = f_0$ where $f_0$ is an odd constant, then, we have
\begin{equation}
{\widetilde \Lambda} = D + \left \lb \xi \(D^3 \ln \xi \) + \xi^2 f_0 
\right\rb D^{-2} \xi^{-1}
\label{ltia}
\end{equation}
which is of the same form as ${\widetilde \Lambda}$.

The second possibility, on the other hand, is more interesting.
Let us choose $\xi$  such that
\begin{equation}
\xi \(D^3 \ln \xi \) + \xi \(\xi^{-1} \psi \int (\phi \xi ) \)^{\prime}=0
\label{second}
\end{equation}
With this choice it follows from (\ref{tdt}) and
(\ref{tpdpt}) that
\begin{equation}
T {\widetilde \Lambda} T^{-1} = D - \xi \(\xi^{-1} \psi  \)^{\prime}
D^{-2} \(\int (\phi \xi )\) \xi^{-1} 
\label{tplpt}
\end{equation}
has the  desired form.

Condition (\ref{second}) can be rewritten as 
(by ignoring one odd integration constant)
\begin{equation}
D \xi + \psi \int (\phi \xi ) =0
\label{seconda}
\end{equation}
Remarkably, this is nothing but a condition ${\widetilde \Lambda} (\xi) =0$.
We will use this fact later.
Define now
\begin{equation}
{\widetilde \Lambda}_2 = T_1 {\widetilde \Lambda}_1 T_1^{-1} = 
T_1 T_0 D T_0^{-1} T_1^{-1}\quad ;\quad
T_0 = \xi_0 \pa \xi^{-1}_0 \quad ;\quad T_1 = \xi_1 \pa \xi^{-1}_1
\label{ltwo}
\end{equation}
One sees that ${\widetilde \Lambda}_1$ is given by the right hand side
of (\ref{tdt}).
The condition (\ref{seconda}) ${\widetilde \Lambda}_1 (\xi_1)=0$ for $\xi_1$ 
has a solution of the form 
$ \xi_1 = {\rm const}\times \pa \ln \xi_0$.
For this solution we find (with ${\rm const} =1$):
\begin{equation}
{\widetilde \Lambda}_2 = T_1 {\widetilde \Lambda}_1 T_1^{-1} = D + 
(\pa \xi_0) \( D^3 \ln (\pa \xi_0)\)  D^{-2} (\pa \xi_0)^{-1} 
\label{ltwoa}
\end{equation}
These observations inspire us to define a chain of the \DB transformations
\begin{equation}
{\widetilde \Lambda}_n = T_{n-1} {\widetilde \Lambda}_{n-1}  
T_{n-1}^{-1} \quad ;\quad T_n = \xi_n \pa \xi^{-1}_n
\label{chain}
\end{equation}
such that the constraint $ {\widetilde \Lambda}_n (\xi_n)=0$
is satisfied at each level.
Correspondingly,
\begin{eqnarray}
{\widetilde \Lambda}_n  &=& D + \psi_n D^{-2} \phi_n
\label{ln}\\
\phi_n &=& (T_{n-1}^{-1} )^{\dag} \cdots (T_{1}^{-1} )^{\dag} ( \xi_0^{-1})
\label{pn}\\
\psi_n &=& T_{n-1} \cdots T_{1}  ( \xi_0 D^3 \ln \xi_0 )
= T_{n-1} \cdots T_{0} ( D \xi_0 )
\label{psin}
\end{eqnarray}
The closed expressions for $\xi_n, \phi_n, \psi_n$ 
\begin{eqnarray}
\xi_n &=& \pa \ln \( \pa^{n-1} \xi_0 \) \label{xin}\\
\phi_n &=& \( \pa^{n-1} \xi_0 \)^{-1}  \label{pna}\\
\psi_n &=& \( \pa^{n-1} \xi_0 \) D^3 \ln \( \pa^{n-1} \xi_0 \)
\label{psina}
\end{eqnarray}
can be verified by inspection.
One also finds  that
$ {\bar \phi}_n = {\widetilde \Lambda}_n^{\dag} (\phi_n )=0$ for all $n$
and therefore the bosonic Lax operator \begin{equation}
{\widetilde \Lambda}_n^2 = \pa + \( \pa^{n-1} \xi_0 \) \(\ln \( \pa^{n-1}
\xi_0 \)\)^{\prime \prime} \pa^{-1} \( \pa^{n-1} \xi_0 \)^{-1}
\label{boslaxn}
\end{equation}
obtained by squaring ${\widetilde \Lambda}_n$ like in (\ref{C_{17}})
will only have bosonic coefficients, while fermionic coefficients do not
appear. 
Similarly, in the s-AKNS equations of motion for these solutions the
quadratic fermionic terms will be absent.
This feature is reminiscent of the soliton solutions found for the 
sAKNS model in [13] using the algebraic dressing method.
\vspace{.5cm}

\begin{large}
\noindent{\bf 6. Conclusion:}
\end{large}

\vspace{.5cm}
As we have argued, the sAKNS hierarchy is clearly an important supersymmetric
integrable hierarchy. In this paper, we have worked out various properties
associated with this hierarchy systematically. We have constructed the
conserved charges (both bosonic and fermionic), the Hamiltonian structures
verifying the Jacobi identity and the recursion operator. We have also shown
that this is a unique hierarchy which allows a description in terms of a
bosonic as well as a fermionic Lax operator. Starting from the fermionic Lax
operator, we have shown how the bosonic conserved charges as well as the zero
curvature formulation can be obtained. The Darboux-B\"acklund transformation
associated with the fermionic Lax operator is also worked out leading to
soliton solutions. There remain, however, some open questions. Namely, we do
not yet know how to obtain the fermionic conserved quantities starting from 
the fermionic Lax operator nor is a direct connection between the bosonic 
and the fermionic Lax operator clear at this point. However, we would like to
emphasize that it is quite worth understanding in detail the description of
this hierarchy in terms of the fermionic Lax operator since it might give some
insight into the Manin-Radul description of the supersymmetric 
KP hierarchy[7].

This work was supported in part by US Department of Energy Grant No.
DE-FG02-84ER40173 and DE-FG02-91ER40685. 
\vspace{.5cm}

\begin{large}
\noindent{\bf References:}
\end{large}

\vspace{.5cm}
\begin{enumerate}
\item{ }V. E. Zakharov and A. B. Shabat, JETP {\bf 34}, 62 (1972).
\item{ }M. J. Ablowitz et al, Phys. Rev. Lett. {\bf 30}, 1262 (1973); {\it
ibid} {\bf 31}, 125 (1973).
\item{ }A. Das, {\it Integrable Models}, World Scientific (1989). 
\item{ }A. P. Fordy and P. P. Kulish, Comm. Math. Phys. {\bf 89}, 427 (1983). 
\item{ }H. Aratyn, J. F. Gomes and A. H. Zimerman, J. Math. Phys. {\bf 36},
3419 (1995).
\item{ }B. A. Kupershmidt, {\it Elements of Super Integrable Systems: basic
techniques and results}, Kluwer Acad. Publ. (1987).
\item{ }Y. Manin and A. O. Radul, Comm. Math. Phys. {\bf 98}, 65 (1985).
\item{ }P. Mathieu, J. Math. Phys. {\bf 29}, 2499 (1988).
\item{ }J. C. Brunelli and A. Das, Phys. Lett. {\bf 337B}, 303 (1994); Int. J.
Mod. Phys. {\bf A10}, 4563 (1995).
\item{ }H. Aratyn and C. Rasinariu, Phys. Lett. {\bf 391B}, 99 (1997).
\item{ }Z. Popowicz, J. Phys. {\bf A29}, 1281 (1996).
\item{ }H. Aratyn, A. Das and C. Rasinariu, Mod. Phys. Lett. A, to be
published.
\item{ }H. Aratyn, A. Das, C. Rasinariu and A. H. Zimerman, Lecture Notes in
Physics, to be published.
\item{ }P. J. Olver, {\it Applications of Lie Groups to Differential
Equations}, Graduate Texts in Mathematics, vol 117, Springer, New York (1986).
\item{ }P. Mathieu, Lett. Math. Phys. {\bf 16}, 199 (1988).
\item{ }J. C. Brunelli and A. Das, Phys. Lett. {\bf 354B}, 307 (1995).
\item{ }K. Becker and M. Becker, Mod. Phys. Lett. {\bf A8}, 1205 (1993); J. M.
Figueroa-O'Farril and S. Stanciu, Phys. Lett. {\bf 316B}, 282 (1993).
\item{ }J. C. Brunelli and A. Das, Phys. Lett. {\bf 409B}, 229 (1997).
\end{enumerate}

\vfil
\eject 

\end{document}